\documentclass[published]{JHEP3} % 10pt is ignored!

\JHEP{00(2002)000}

\JHEPspecialurl{http://jhep.sissa.it/JOURNAL/JHEP3.tar.gz}

\usepackage{epsfig,multicol,bbm}

\newcommand\fverb{\setbox\pippobox=\hbox\bgroup\verb}
\newcommand\fverbdo{\egroup\medskip\noindent%
                        \fbox{\unhbox\pippobox}\ }
\newcommand\fverbit{\egroup\item[\fbox{\unhbox\pippobox}]}
\newbox\pippobox
\def\a{\alpha}
\def\b{\beta}
\def\g{\gamma}

\def\D{\Delta}
\def\e{\epsilon}

\def\t{\theta}

\def\be{\begin{equation}}
\def\ee{\end{equation}}
\def\ba{\begin{eqnarray}}
\def\ea{\end{eqnarray}}

\newcommand{\rpar}{\stackrel{\leftarrow}{\partial}}
\newcommand{\lpar}{\stackrel{\rightarrow}{\partial}}

\newcommand{\no}{\nonumber}

\title{Odd Nambu bracket on Grassmann algebra}

\author{Dmitrij V. Soroka and Vyacheslav A. Soroka\\
        Kharkov Institute of Physics and Technology\\
        1 Akademicheskaya St., 61108 Kharkov, Ukraine\\
        E-mail: \email{dsoroka@kipt.kharkov.ua}, \email{vsoroka@kipt.kharkov.ua}}
\received{}             %%
\revised{}
\accepted{}             %% These are for published papers.

\preprint{\hepth{9912999}}      % OR: \preprint{Aaaa/Mm/Yy\\Aaa-aa/Nnnnnn}
\abstract{The Grassmann-odd Nambu bracket on the Grassmann algebra is 
proposed.}


\dedicated{Dedicated to Valentina Erofeyevna Korol'.\\}

\begin{document} 

%\maketitle  IS IGNORED %%%%%%%%%%%

\section{Introduction}

In the paper \cite{s1} the linear odd bracket corresponding to the $SO(3)$ 
group and realized solely on the Grassmann variables $\t_\a$ has been 
introduced (see also \cite{s2,s3,ss1,ss2})
\ba\label{1.1}
\{A,B\}_1=A\rpar_{\t_\a}\e_{\a\b\g}\t_\g\lpar_{\t_\b}B,\quad(\a,\b,\g=1,2,3),
\ea
where $\e_{\a\b\g}$ is the completely antisymmetric Levi-Civita tensor,
$\rpar$ and $\lpar$ are the right and left derivatives and 
$\partial_{\t_\a} \equiv {\partial \over {\partial \t_\a}}$. It was constructed
for this bracket three Grassmann-odd nilpotent $\D$-like differential operators
of the first, second and third orders with respect to the Grassmann derivatives
\ba\label{1.2}
\D_{+1}=\e_{\a\b\g}\t_\a\t_\b\partial_{\t_\g},
\ea
\ba\label{1.3}
\D_{-1}={1\over2}\e_{\a\b\g}\t_\a\partial_{\t_\b}\partial_{\t_\g},
\ea
\ba\label{1.4}
\D_{-3}={1\over3!}\e_{\a\b\g}\partial_{\t_\a}\partial_{\t_\b}\partial_{\t_\g}.
\ea

The operator $\D_{+1}$ is proportional to the second term in a BRST charge
\ba
Q=\t_\a G_\a-{1\over2}\e_{\a\b\g}\t_\a\t_\b\partial_{\t_\g},\no
\ea
where $\t_\a$ and $\partial_{\t_\a}$ represent the operators for the ghosts 
and ghost momenta respectively and $G_\a$ are generators of the $SO(3)$ group. 
The operator $\D_{-1}$ related to the divergence of the vector field 
$\{\t_\a,A\}_1$
\ba\label{1.5}
\D_{-1}A=-{1\over2}\partial_{\t_\a}\{\t_\a,A\}_1
\ea
is proportional to the true $\D$-operator for the odd bracket (\ref{1.1}).
The operator $\D_{-1}$ (\ref{1.5}) determines the linear odd bracket 
(\ref{1.1}) as a deviation of the Leibniz rule under the usual multiplication
\begin{eqnarray}
\Delta_{-1}(A\cdot B)=(\Delta_{-1}A)\cdot B
+(-1)^{p(A)}A\cdot\Delta_{-1}B
+(-1)^{p(A)}\{A,B\}_1\ .\nonumber
\end{eqnarray}
and simultaneously satisfies the Leibniz rule with respect to the linear odd 
bracket composition
\ba
\D_{-1}(\{A,B\}_1)=\{\D_{-1}A,B\}_1+(-1)^{p(A)+1}\{A,\D_{-1}B\}_1,\no
\ea
where $p(A)$ is a Grassmann parity of the quantity $A$.

In the present paper we show that the operator $\D_{-3}$ (\ref{1.4}) is related
with the Grassmann-odd Nambu bracket \cite{n} on the Grassmann algebra.

\section{Odd Nambu bracket}

By applying the operator $\D_{-3}$ (\ref{1.4}) to the usual product of two 
quantities $A$ and $B$, we obtain
\begin{eqnarray}\label{2.1}
\Delta_{-3}(A\cdot B)=(\Delta_{-3}A)\cdot B
+(-1)^{p(A)}A\cdot\Delta_{-3}B
+(-1)^{p(A)}\D(A,B),
\end{eqnarray}
where the quantity $\D(A,B)$ is
\ba\label{2.2}
\D(A,B)={1\over2}\e_{\a\b\g}\left[\left(\partial_{\t_\a}\partial_{\t_\b}A
\right)\partial_{\t_\g}B+(-1)^{p(A)}\left(\partial_{\t_\a}A\right)
\partial_{\t_\b}\partial_{\t_\g}B\right].
\ea

By acting with the operator $\D_{-3}$ on the usual product of three quantities
$A$, $B$ and $C$, we come to the following relation:
\ba\label{2.3}
\D_{-3}(A\cdot B\cdot C)&=&\left(\D_{-3}A\right)\cdot B\cdot C+
(-1)^{p(A)}A\cdot \left(\D_{-3}B\right)\cdot C+
(-1)^{p(A)+p(B)}A\cdot B\cdot\D_{-3}C\cr&+&(-1)^{p(A)}\D(A,B)C+
(-1)^{p(A)p(B)+p(A)+p(B)}B\D(A,C)\cr&+&(-1)^{p(A)+p(B)}A\D(B,C)+
(-1)^{p(B)}\{A,B,C\}_1,
\ea
where the last term is the Grassmann-odd Nambu bracket 
\ba\label{2.4}
\{A,B,C\}_1=\e_{\a\b\g}\partial_{\t_\a}A\partial_{\t_\b}B\partial_{\t_\g}C
\ea
on the Grassmann algebra.

The divergence of the Nambu bracket (\ref{2.4}) is related with the quantity
$\D(A,B)$ (\ref{2.2}) (see also \cite{sak})
\ba\label{2.5}
\D(A,B)={1\over2}\partial_{\t_\a}\{\t_\a,A,B\}_1,
\ea
whereas the multiplication on the Grassmann variable $\t_\a$ gives the linear
odd bracket (\ref{1.1})
\ba\label{2.6}
\{A,B\}_1=\t_\a\{\t_\a,A,B\}_1.
\ea
Note also the following relation between the operator $\D_{-3}$ (\ref{1.4})
and odd Nambu bracket (\ref{2.4}):
\ba\label{2.7}
\D_{-3}A={1\over3!}\partial_{\t_\a}\partial_{\t_\b}\{\t_\a,\t_\b,A\}_1.
\ea
For the operator $\D_{-3}$ (\ref{1.4}) there exists the following ``Leibniz
rule'' with respect to the odd Nambu bracket composition:
\ba\label{2.8}
\D_{-3}(\{A,B,C\}_1)=&-&\{\D_{-3}A,B,C\}_1+(-1)^{p(A)}\{A,\D_{-3}B,C\}_1\cr
&-&(-1)^{p(A)+p(B)}\{A,B,\D_{-3}C\}_1\cr&+&\e_{\a\b\g}\Bigl[(-1)^{p(A)+1}
\D\left(\partial_{\t_\a}A,\partial_{\t_\b}B\right)\partial_{\t_\g}C\cr
&+&(-1)^{p(A)+p(B)}\partial_{\t_\a}A\D\left(\partial_{\t_\b}B,
\partial_{\t_\g}C\right)\cr&+&(-1)^{p(A)p(B)+1}\partial_{\t_\b}B
\D\left(\partial_{\t_\a}A,\partial_{\t_\g}C\right)\Bigr].
\ea

It follows from the expression (\ref{2.4}) the Grassmann parity
\ba\label{2.9}
p(\{A,B,C\}_1)=p(A)+p(B)+p(C)+1\pmod2,
\ea
symmetry properties
\ba\label{2.10}
\{A,B,C\}_1=-(-1)^{[p(A)+1][p(B)+1]}\{B,A,C\}_1
=-(-1)^{[p(B)+1][p(C)+1]}\{A,C,B\}_1
\ea
and Jacobi type identity for the odd Nambu bracket
\ba\label{2.11}
\{\{A,B,C\}_1,D,E\}_1&+&(-1)^{[p(A)+1][p(B)+p(C)+p(D)+p(E)]}
\{\{B,C,D\}_1,E,A\}_1\cr&+&(-1)^{[p(A)+p(B)][p(C)+p(D)+p(E)+1]}
\{\{C,D,E\}_1,A,B\}_1\cr&+&(-1)^{[p(D)+p(E)][p(A)+p(B)+p(C)+1]}
\{\{D,E,A\}_1,B,C\}_1\cr&+&(-1)^{[p(E)+1][p(A)+p(B)+p(C)+p(D)]}
\{\{E,A,B\}_1,C,D\}_1\cr&+&(-1)^{[p(D)+p(E)][p(A)+p(B)+p(C)+1]+p(B)[p(A)+1]
+p(A)}\cr&\times&\{\{D,E,B\}_1,A,C\}_1\cr&+&(-1)^{[p(D)+1][p(A)+p(B)+p(C)]
+p(D)}
\{\{D,A,B\}_1,C,E\}_1\cr&+&(-1)^{[p(A)+1][p(B)+p(C)+p(D)+p(E)]+[p(D)+1]p(E)
+p(D)}\cr&\times&\{\{B,C,E\}_1,D,A\}_1\cr&+&(-1)^{[p(B)+1][p(C)+p(D)+p(E)]
+p(B)}
\{\{A,C,D\}_1,E,B\}_1\cr&+&(-1)^{[p(B)+1][p(C)+p(D)+p(E)+1]+[p(D)+1][p(E)+1]}
\cr&\times&\{\{A,C,E\}_1,D,B\}_1=0.
\ea
Note that the structure of (\ref{2.11}) is different from the one for the
fundamental identity \cite{t}.

\section{Conclusion}

Thus, we constructed the Grassmann-odd Nambu bracket on the Grassmann
algebra. The main properties of this bracket are also given.

\acknowledgments

One of the authors (V.A.S.) thanks the administration of the Office of 
Associate and Federation Schemes of the Abdus Salam ICTP for the kind 
hospitality at Trieste where this work has been performed.


\begin{thebibliography}{999}

\bibitem{s1}V.A. Soroka, \emph{ Degenerate odd Poisson bracket on Grassmann 
variables}, \pan{63}{2000}{915}; \hepth{9811223}. 
\bibitem{s2}V.A. Soroka, \emph{ Linear odd Poisson bracket on Grassmann 
variables}, \plb{451}{1999}{349}; \hepth{9811252}. 
\bibitem{s3}V.A. Soroka, \emph{ Supersymmetry and the odd Poisson bracket},
\npps{101}{2001}{26}; \hepth{0204018}. 
\bibitem{ss1}D.V. Soroka and V.A. Soroka, \emph{Lie-Poisson (Kirillov) odd 
bracket on Grassmann algebra},
Proceedings of the XXIII International Colloquium
"Group Theoretical Methods in Physics" (GROUP 23, July 31 -- August 5,
JINR, Dubna, Russia, 2000) Vol. 1, pp. 106-111, Dubna, 2002.
\bibitem{ss2}D.V. Soroka and V.A. Soroka, \emph{ Poisson-Lie odd bracket on 
Grassmann algebra}, \newjournal{SIGMA\ }{SIGMA}{2}{2006}{Paper 036}; 
\hepth{0603148}. 
\bibitem{n}Y. Nambu, \emph{Generalized Hamiltonian dynamics}, 
\prd{7}{1973}{2405}.
\bibitem{sak}M. Sakakibara, \emph{Notes on the super Nambu bracket}, 
\ptp{109}{2003}{305}; math-ph/0208040.
\bibitem{t}L. Takhtajan, \emph{On foundation of the generalized Nambu 
mechanics}, \cmp{160}{1994}{295}.

\end{thebibliography}
\end{document}